\documentclass[a4paper,dvips,12pt]{article}
\usepackage{amsmath,amssymb,exscale}   
\usepackage{array,multicol}
\usepackage{afterpage,float,flafter}   
\usepackage{epsfig,rotating,pifont,fancybox}   
\makeatletter   
\@addtoreset{equation}{section}   
\makeatother   
\newcommand{\bea}{\begin{eqnarray}}   
\newcommand{\eea}{\end{eqnarray}}   
   
\newcommand{\NPB}[3]{\emph{ Nucl.~Phys.} \textbf{B#1} (#2) #3}   
\newcommand{\PLB}[3]{\emph{ Phys.~Lett.} \textbf{B#1} (#2) #3}   
\newcommand{\PRD}[3]{\emph{ Phys.~Rev.} \textbf{D#1} (#2) #3}   
\newcommand{\PRL}[3]{\emph{ Phys.~Rev.~Lett.} \textbf{#1} (#2) #3}   
\newcommand{\ZPC}[3]{\emph{ Z.~Phys.} \textbf{C#1} (#2) #3}   
\newcommand{\PTP}[3]{\emph{ Prog.~Theor.~Phys.} \textbf{#1}  (#2) #3}

\newcommand{\JHEP}[3]{\emph{ JHEP} \textbf{#1} (#2) #3}

\title{   
\vspace*{-1.2cm}   
\begin{flushright}   
\normalsize{      
LPTENS-00/18\\
IEM-FT-202/00\\
IFT-UAM/CSIC-00-17\\   
\texttt{hep-ph/0004124}}\\ 
\end{flushright}    
\vspace{1cm}
\Large\textbf{The lightest Higgs mass in supersymmetric models with
extra dimensions~\footnote{Work 
supported in part by EU under TMR contract ERBFMRX-CT96-0045, and
by CICYT (Spain) under contract AEN98-0816.}}
\author{\large\textbf
{A.~Delgado$^a$ and M.~Quir{\'o}s$^{a,b}$}\\ \\ 
$^a$\emph{Instituto de Estructura de la Materia (CSIC),}\\
\emph{Serrano 123, E-28006, Madrid, Spain}\\
$^b$\emph{Laboratoire de Physique Th\'eorique (ENS), 24 rue Lhomond,}\\
\emph{F-75231, Paris, France~\footnote{Unit\'e mixte du CNRS et
de l'ENS, UMR 8549.}}}}
\date{}   
\begin{document}
\maketitle


\begin{abstract}
In the four-dimensional supersymmetric standard model extended with 
gauge singlets the lightest Higgs boson mass, $M_H$, has an important
contribution proportional to the squared of the
superpotential coupling $\lambda$ of singlets to Higgs fields, 
$\lambda\, SH_1\cdot H_2$. The requirement of perturbativity up to the
unification scale yields an upper bound on $M_H\sim
140$ GeV. In extensions to theories with (longitudinal) 
extra dimensions at the TeV where such
coupling exists and massive Kaluza-Klein states fall into $N=2$
representations, if either of the Higgs or singlet fields live in the
bulk of the extra dimensions, the $\beta$-function of $\lambda$ is
suppressed due to the absence of anomalous dimension of
hypermultiplets to leading order. This implies a slower running of
$\lambda$ and an enhancement of its low energy value. 
The $M_H$ upper bound increases to values $M_H\lesssim 165$ GeV. 
\end{abstract}
\vspace{1.5cm}   
   
\begin{flushleft}   
April 2000 \\   
\end{flushleft}   
\newpage

\noindent
LEP unification of the gauge couplings in the minimal supersymmetric
extension of the Standard Model (MSSM), possibly including gauge
singlets (NMSSM), is the only ``hint'' of new physics at present accelerators.
In fact, running the gauge couplings from their LEP values ($M_Z$) to
high energies with the (N)MSSM four-dimensional (4D) $\beta$-functions 
one finds a remarkable unification of the $SU(3)\times
SU(2)\times U(1)$ couplings at a scale $M_{U}^{4D}\simeq 2\times 10^{16}$ GeV.
Furthermore, the existence of a light Higgs with a tree-level upper bound
(independent of the values of the soft masses of all
other scalar partners) on its mass, $M_H$, is one of the most solid, and model
independent, predictions of supersymmetric theories. In particular for
the MSSM there exists the tree-level bound $M_H\leq M_Z |\cos 2\beta|$
which, after inclusion of radiative corrections yields, for the
experimental value of the top-quark mass, an absolute upper bound
$M_H\lesssim 125$ GeV~\cite{old,carena}. For the NMSSM, assuming a 
superpotential coupling of gauge singlets $\vec{S}$ as
\begin{equation}
\label{superp}
W=\vec{\lambda}\, \vec{S}H_1\cdot H_2 \ ,
\end{equation}
one finds the tree level upper bound for the lightest Higgs mass:
\begin{equation}
\label{cota}
M_H^2\leq \left( \cos^2 2\beta+\frac{2 \vec{\lambda}^2
\cos^2\theta_W}{g_2^2} \sin^2 2\beta\right)M_Z^2\ .
\end{equation}
Imposing perturbativity up to $M_{U}^{4D}$, and including radiative
corrections, an upper bound for the lightest Higgs mass as
$M_H\lesssim 140$ GeV \cite{espinosa1} is found.

Being the above bounds for the Higgs mass rather restrictive, the
corresponding models could be ruled out at the next run of proton
colliders (TeVatron, LHC) if the Higgs is not seen in the corresponding
ranges, irrespectively of any direct search for supersymmetric
partners. A way of increasing the absolute upper bound on the lightest
Higgs mass in the NMSSM, without spoiling the unification
properties of the gauge couplings, was proposed 
by introducing extra lepton and quark
fields in $SU(5)$ representations, plus complex conjugates to avoid 
anomalies appearance. These models neither spoil unification nor change the
unification scale, but they increase the value of the unification
coupling, $\alpha_{U}^{4D}$, moving it towards the non-perturbative regime. 
Increasing the gauge couplings all the way between $M_Z$ and
$M_{U}^{4D}$ slows down the rate of running of $\vec{\lambda}$, which can
then acquire larger initial values while remaining perturbative up to
$M_{U}^{4D}$. Obviously this effect increases the upper bound on the
lightest Higgs mass. By adding extra matter at scales $\gtrsim$ 1 TeV in
the representation $\mathbf{5}+\bar\mathbf{5}$ it was shown that 
$M_H\lesssim 155$ GeV \cite{pomarol1,espinosa2}, that can be
considered as the absolute upper bound.

In this Letter we will propose another possible way of increasing the absolute
bound on $M_H$, without neither spoiling the unification properties of the
MSSM nor moving $\alpha_{U}$ to the non-perturbative
regime. It is based on the existence of extra dimensions at the
TeV scale, where the gauge bosons and some of the Higgs and/or matter
fields live. By using this genuine extra-dimensional mechanism we will
see that the upper bound on $\vec\lambda$ (i.e. on $M_H$) can be considerably
enhanced with respect to the NMSSM case. In the simplest example that
we will explicitly work out, based on one-extra dimension compactified
on the orbifold $S^1/\mathbb{Z}_2$, the absolute upper bound is increased to 
$M_H\lesssim 165$ GeV. Finally we will prove how robust this result is versus
the number of extra dimensions.

The existence of large (TeV) extra dimensions feeling gauge
interactions has been shown in a large class of string
theories, including type I and type IIB vacua. The possibility of
lowering the compactification scale $M_c\equiv 1/R$ to the TeV 
was first proposed in Ref.~\cite{ignatios}, 
while the option of lowering the string (quantum gravity)
scale, $M_s$, was suggested in Ref.~\cite{likken}. Having the fundamental scale
close to the TeV implies that some ($\delta$) extra dimensions ``parallel'' to
the 4D space can appear at the TeV while other, ``transverse'', ones
(where gravity lives) can be much larger 
(sub-millimeter dimensions)~\cite{review}. 

If $M_s\gg M_c$ there are threshold corrections for the
 couplings at $M_c$, arising from loops of Kaluza-Klein (KK)
excitations, that can be interpreted as a 
power-law running~\cite{veneziano} and can
lead to unification of gauge couplings at some scale $M_U$ that we
will identify with the string scale~\cite{dienes}~\footnote{We also expect in
general (model dependent) string threshold corrections. However in the
regime $M_s\gg M_c$ they are subleading compared to the
contribution from the massive KK modes and will not be considered 
here.}. Depending on the fields living
in the bulk of the extra dimensions (apart from the gauge fields that
we always assume to live in it), we can have unification of gauge
couplings, but at a much lower scale than $M_U^{4D}$. 

In this work we will first focus on the case of just one
``parallel'' dimension ($\delta=1$) at the TeV scale, where gauge
bosons and some of the Higgs and matter multiplets live. As we will
see the results are easily generalized to the case $\delta>1$.
We consider a 5D theory compactified on
$S^1/\mathbb{Z}_2$~\cite{mp,dpq}. Vector fields in the bulk are in
$N=2$ vector multiplets, 
$\mathbb{V}=(A_\mu,\lambda_1;\Phi,\lambda_2)$~\footnote{Where $i=1,2$
transform as $SU(2)_R$ indices and the complex scalar $\Phi$ is
defined as, $\Phi\equiv A_5+i\Sigma$.}, 
while matter fields in the bulk are arranged in $N=2$
hypermultiplets, $\mathbb{H}=(z_1,\psi_R;z_2,\psi_L)$. $\mathbb{Z}_2$
is acting as a parity in the fifth dimension, $x_5\to -x_5$ and, with
an appropriate lifting to spinors and $SU(2)_R$ indices, we can
decompose $\mathbb{V}$ and $\mathbb{H}$ into \emph{even},
$(A_\mu,\lambda_1)$ and $(z_1,\psi_R)$, and \emph{odd},
$(\Phi,\lambda_2)$ and $(z_2,\psi_L)$, $N=1$ superfields.
After the $\mathbb{Z}_2$ projection the only
surviving zero modes are in $N=1$ vector and chiral multiplets. If the
chiral multiplet $(z_1,\psi_R)$ is not in a real representation of the
gauge group, a multiplet of opposite chirality localized in the 4D
boundary $(\widetilde{z},\widetilde{\psi}_L)$ can be introduced 
to cancel anomalies. In
order to do that $\mathbb{Z}_2$ must have a further action on
the boundary under which all boundary states are 
odd~\footnote{We thank C. Kounnas for interesting discussions 
about this point.}. 
This action can be defined as $(-1)^{\varepsilon_i}$ for the
chiral multiplet $X_i$, such that $\varepsilon_i=1$ (0) for $X_i$
living in the boundary (bulk). The MSSM is then made up of zero
modes of fields living in the 5D
bulk and chiral $N=1$ multiplets in the 4D boundary, at 
localized points of the bulk. The Higgs and matter sector
($H_1,\ H_2,\ Q,\ U,\ D,\ L,\,E$) will then be allowed to live, either in
the bulk as independent hypermultiplets or at the boundary as chiral
multiplets. In either case, after the $\mathbb{Z}_2$ projection, they
lead to the supermultiplet structure of the MSSM.

The general MSSM unification condition in the presence of extra dimensions
was obtained in Ref.~\cite{unif5D} as:
\begin{equation}
\label{unif}
2\, N_E+5\, N_U-7\, N_Q+3\left(N_D-N_L\right)
-3\left(N_{H_1}+N_{H_2}\right)+2
-32\, T_0-20\, T_1=0 \ ,
\end{equation}
where $N_X$ ($X=H_1,\ H_2,\ Q,\ U,\ D,\ L,\,E$) is the number of
$X$ hypermultiplets. We have considered the possibility of $T_Y$ pairs
of hypermultiplets in the bulk 
which transform as $(\mathbf{1},\mathbf{3})_{|Y|}$ under 
$SU(3)\times SU(2)\times U(1)$.
The hypermultiplets which are not included in the
usual generations, or transform as real representations of the gauge
group, must be considered in pairs to cancel the anomalies
of the zero modes after the $\mathbb{Z}_2$ projection. We assume in 
Eq.~(\ref{unif}) all
gauge bosons living in the bulk, which is the natural possibility to
achieve unification of gauge couplings~\cite{imr}. The relationship between the
unification scales $M_U$ and $M_U^{4D}$ is given by,
\begin{equation}
\label{escala}
\frac{M_U}{M_c}= \rho\, \log\frac{M_U^{4D}}{M_c} \ ,
\end{equation}
where 
$$\rho=14/[10+3 N_E - N_L-\left(N_{H_1}+N_{H_2}\right)
-7 N_Q+4 N_U+ N_D-10 T_0-2 T_1]\ .$$ 
The meaning of Eq.~(\ref{escala}) is
clear: in models with power-law running (linear in the case of a single extra
dimension) the evolution of gauge couplings is much
faster than in models with logarithmic running. The former ones run less
(left-hand side of (\ref{escala})) than the latter ones
(right-hand side of (\ref{escala})) and, as a consequence, unify earlier. 
In models with $\rho=1$ the different evolution rates are exactly
compensated by different unification scales. Therefore 
MSSM-like unification  is more
natural in models with $\rho=1$, as we will require in this 
work~\footnote{Of course one can find unifying models satisfying (\ref{unif})
with $\rho<1$. They will not be considered in detail in this
work since they require more extra matter and easily conflict the original
condition that $M_U\gg M_c$. In fact the constraint $\rho=1$ is more
accute in models with $\delta>1$.
We will comment later on about these possibilities.}.

In what follows we will be concerned with higher dimensional models
that unify as the MSSM, Eq.~(\ref{unif}), and where there are
superpotential couplings on the 4D boundary, as in Eq.~(\ref{superp}),
that can increase the lightest Higgs tree-level mass as in
Eq.~(\ref{cota}). In general, for the existence of a superpotential
coupling as $f^{i_1\, \dots\, i_N}X_{i_1}\, \cdots X_{i_N}$, 
the orbifold selection rule 
\begin{equation}
(-1)^{\varepsilon_{i_1}+\dots+\varepsilon_{i_N}}=1
\label{condition}
\end{equation}
is required~\cite{zurab1,dpq2}. This condition implies, for
renormalizable couplings, that 
$\varepsilon_i+\varepsilon_j+\varepsilon_k=2$~\footnote{The possibility
$\varepsilon_i+\varepsilon_j+\varepsilon_k=0$ is not excluded by the
orbifold group action but it is suppressed, with respect to the
former one, by an extra factor $\sim 1/(M_s R)\ll 1$~\cite{dpq2}.}. 
Electroweak precision measurements~\cite{gatto,dpq2}, as well as direct
searches~\cite{direct}, 
constrain, due to the presence of KK excitations of Standard 
Model gauge bosons, the value of the compactification scale $M_c$ in
each case, providing bounds which depend on the chosen values of the
parameters $\varepsilon_i$ and $\tan\beta$~\cite{dpq2}. Those bounds will be
taken into account in the subsequent analysis. 

In the considered models 
the gauge bosons are living in the
bulk. They can be used to break supersymmetry by their
interactions with the hidden wall~\cite{gmsb}, or by the compactification
mechanism~\cite{dpq}. In either case they induce a gauge mediated supersymmetry
breaking (GMSB) scenario. 
Our subsequent analysis of the lightest Higgs mass is not
sensitive to the details of supersymmetry breaking, or spectrum, that 
will consequently
be of no concern in this paper. On the other hand the Higgs fields
$H_1$ and $H_2$ can, either of them, live in the bulk of the extra
dimension or in the 4D boundary. In each case the consequences for the
Higgs mass will be different, as we will see.

If both $H_1$ and $H_2$ are living in the bulk 
($N_{H_1}=N_{H_2}=1$) and superpotential
Yukawa couplings are assumed for leptons and quarks,
Eq.~(\ref{condition}) can be satisfied provided that all three generation 
matter multiplets live in the 4D boundary. Unification as in the
MSSM, with $\rho=1$, can be easily performed by introducing two extra
$E$-like hypermultiplets at the compactification scale, $\Delta
N_E=2$~\footnote{A supersymmetric mass term for pairs of
hypermultiplets can be easily obtained by a supersymmetry preserving
compactification~\cite{dpq}.}. 
This model was introduced in Ref.~\cite{zurab} and it is 
contained in condition (\ref{unif}). In this case no coupling as in 
Eq.~(\ref{superp}) is consistent with Eq.~(\ref{condition}), neither
for singlets in the bulk or in the 4D boundary~\footnote{As 
previously indicated the coupling (\ref{superp}) is consistent with
the orbifold action (\ref{condition}) for singlets in the bulk, 
but this coupling is
expected to be highly suppressed and its contribution in (\ref{cota})
expected to be negligible.}. The second term in Eq.~(\ref{cota}) is
zero and the lightest Higgs mass bound coincides with that of the
MSSM, i.e. the first term in Eq.~(\ref{cota}). 

If the $H_1$ and $H_2$ Higgs fields live in the 4D boundary
($N_{H_1}=N_{H_2}=0$) and Yukawa couplings for leptons and quarks are
assumed, it follows from Eq.~(\ref{condition}) that some matter multiplets
must live in the 4D boundary while others must propagate in the 5D bulk. A
discussion of the different possibilities was performed in
Ref.~\cite{dpq2}, where lower bounds for the different cases were found
from electroweak precision measurements. There exists a minimal model with
MSSM-like unification and $\rho=1$: right-handed quarks and
leptons live in the bulk, left-handed quarks and leptons in the 4D
boundary, $N_U=N_D=N_E=3$, and there is also a pair of zero-hypercharge
triplets at the scale $M_c$ which live in the bulk, $T_0=1$. The
presence of triplets at $M_c$ plays no role for the lightest Higgs
mass bound, since they are introduced 
merely for the purpose of unification. Had we not introduced them the
model would not unify but the corresponding Higgs mass bounds 
would be only slightly modified. In this case $\vec{S}$-singlet(s) 
in the bulk  can couple to  $H_1\cdot H_2$ with the coupling
$\vec\lambda$ in the superpotential (\ref{superp}) and the 
tree-level Higgs mass bound (\ref{cota}) applies. 

As in 4D models, in order to evaluate the absolute upper bound on the lightest
Higgs mass we need to compute the upper bound on $\vec\lambda$ at low
energy by imposing perturbativity of the theory up to the unification
scale $M_U$. The relevant Yukawa couplings are in this case $h_t$, $h_b$
(the top and bottom quark Yukawa couplings) and $\lambda$ (we will
assume one singlet or orientate $\vec\lambda$ along one particular
direction). For $M_Z\leq\mu\leq M_{\rm SUSY}$~\footnote{We are
assuming a common supersymmetric mass for all Standard Model
superpartners $M_{\rm SUSY}$.} 
the beta functions for 
gauge and Yukawa couplings are those of the Standard Model, and for 
$M_{\rm SUSY}\leq\mu\leq M_c$ those of the MSSM. For $\mu\geq M_c$ the
extra dimension (KK-modes) is felt by the couplings and 
the one-loop $\beta$-functions for Yukawa couplings can be written as:
\begin{align}
\label{RGE}
16\pi^2\, \beta_{\lambda}=&\left(\gamma^{S}_{\ S}+\gamma^{H_1}_{\ H_1}+ 
\gamma^{H_2}_{\ H_2}\right)\, \lambda\nonumber\\
16\pi^2\, \beta_{h_t}=&\left(\gamma^{Q}_{\ Q}+\gamma^{U}_{\ U}+ 
\gamma^{H_2}_{\ H_2}\right)\, h_t \nonumber\\
16\pi^2\, \beta_{h_b}=&\left(\gamma^{Q}_{\ Q}+\gamma^{D}_{\ D}+ 
\gamma^{H_1}_{\ H_1}\right)\, h_b \ .
\end{align}
The one-loop wave-function renormalization is given by~\footnote{We use
the renormalization scheme of Ref.~\cite{dienes}, where the factor $e^t$ in
(\ref{wavedef}) is accompanied by $f_{\delta}=X_{\delta}/N$, 
with $\delta=1$, where
$X_{1}=2$ and the factor $N=2$ is coming from the $\mathbb{Z}_2$
orbifold action. We will prove that the physical Higgs mass bound 
is indeed independent of the chosen renormalization scheme.}
\begin{equation}
\label{wavedef}
\gamma^{X}_{\ X}\equiv \left\{e^{t}\,
\varepsilon_{X}+\left(1-\varepsilon_{X}\right)
\right\}\widetilde{\gamma}^{X}_{\ X}\ ,
\end{equation}
where $t$ is related to the RGE scale $\mu$ by $t\equiv
\log(\mu/\mu_0)$, and~\cite{zoupanos,manel} 
\begin{align}
\label{wave}
\widetilde{\gamma}^{S}_{\ S}=&\ 2\lambda^2 \nonumber\\
\widetilde{\gamma}^{H_1}_{\ H_1}=&-\frac{3}{2}g_2^2-\frac{3}{10} g_1^2
+3 h_b^2+\lambda^2 \nonumber\\
\widetilde{\gamma}^{H_2}_{\ H_2}=&-\frac{3}{2}g_2^2-\frac{3}{10} g_1^2
+3 h_t^2+\lambda^2 \nonumber\\
\widetilde{\gamma}^{Q}_{\ Q}=&\  h_t^2+h_b^2-\left(\frac{1}{30} g_1^2
+\frac{3}{2} g_2^2+\frac{8}{3} g_3^2 \right)\nonumber\\
\widetilde{\gamma}^{U}_{\ U}=&\ 2\, h_t^2-\left(\frac{8}{15} g_1^2
+\frac{8}{3} g_3^2 \right)\nonumber\\
\widetilde{\gamma}^{D}_{\ D}=&\ 2\, h_b^2-\left(\frac{2}{15} g_1^2
+\frac{8}{3} g_3^2 \right) \ .
\end{align} 
Note that for $X$-supermultiplets in the 4D boundary,
$\varepsilon_X=1$, the contribution of KK-modes to the anomalous
dimension is the same as that of the zero modes, what explains the
power-law behaviour in (\ref{wavedef}), while for $X$-fields in the
bulk, with KK-modes arranged into $N=2$ multiplets, there is no
anomalous dimension to leading order 
and the only contribution turns out to be the same as for
$\mu\leq M_c$. The $\beta$-functions for gauge couplings in the
present model are~\cite{unif5D}
\begin{align}
\label{RGE2}
16\pi^2\, \beta_{g_1}=& \left(\frac{48}{5}\, e^t\, -3\right) g_1^3
\nonumber\\
16\pi^2\, \beta_{g_2}=& \left(4\, e^t\, -3\right) g_2^3
\nonumber\\
16\pi^2\, \beta_{g_3}=& -3\, g_1^3 \ .
\end{align}

A glance at Eqs.~(\ref{RGE})-(\ref{wave}) shows the main difference
with respect to the 4D case, where only zero modes are present and the running
goes up to $M_{U}^{4D}$. The
4D $\beta$-function for $\lambda$ behaves as
$16\pi^2\beta^{4D}_\lambda=b_{\lambda}\lambda^3 \,+\cdots$,
with $b_{\lambda}=4$. Were $\beta_\lambda$ to
behave similarly with respect to $\lambda$, but with a power-law
behaviour, the bounds on $\lambda$ we had obtained would have been similar to
those in the 4D case since, as happens to the gauge couplings, 
the shortening
in the running scale would have been compensated by the faster (power-law)
running of $\lambda$. However, since the
$S$-hypermultiplet lives in the bulk, and has excited modes, 
the $\beta$-function for $\lambda$ behaves as
$16\pi^2\beta_\lambda=\tilde{b}_\lambda \,\lambda^3\, e^t+\cdots$, 
with $\tilde{b}_\lambda=2$, and the running is
slowed down, leading to an enhanced value of $\lambda$ at low
energy. In fact in the crude approximation of neglecting all terms in
$\beta_\lambda$, except the $\lambda^3$ term,
 one obtains an approximate relation
between the low energy values of $\lambda$ and $\lambda_{4D}$ as,
$\lambda^2\simeq r\,\lambda^2_{4D}$ with 
\begin{equation}
\label{erre}
r=\frac{b_{\lambda}}{ \rho\, \tilde{b}_\lambda}\ ,
\end{equation}
where we have crudely taken
$\lambda^{-1}(M_U)=\lambda^{-1}_{4D}(M_U^{4D})=0$ which provides an
overestimate of $M_H$ as we will see, but accurate enough for our
purposes here. 
This leads to an approximate value of the upper bound on $M_H$ in terms of
$M_H^{4D}$, the upper bound in the 4D theory, and $M_H^{MSSM}$, the
value of the Higgs mass in the MSSM, as
\begin{equation}
\label{upperapp}
M_H^2\simeq r\, \left(M_H^{4D}\right)^2-(r-1)\,
\left(M_H^{MSSM}\right)^2 \ .
\end{equation}
In our case we have $r=2$ and using the values of $M_H^{4D}$
($\simeq 140$
GeV) and $M_H^{MSSM}$ ($\simeq 105$ GeV), for $\tan\beta\simeq 2$, $M_{\rm
SUSY}\simeq 1$ TeV and maximal mixing in the stop sector, we find 
$M_H\simeq 165$ GeV, which is, as anticipated, a little overestimate
($\sim 2\%$) of the numerical value we will find ($\simeq 162$ GeV).

In the numerical calculation we
must allow for boundary conditions consistent with perturbativity in
the whole range $M_c\leq \mu\leq M_U$. Including higher loop
corrections, the $\beta$-functions for all the parameters,
e.g. $\lambda$, can be decomposed as
$\beta_\lambda=\beta^{(1)}_\lambda+ \beta^{(2)}_\lambda+\cdots$. We
will consider the theory is perturbative at a given scale $\mu$ 
if $|\beta^{(2)}_\lambda(\mu)|\lesssim
|\beta^{(1)}_\lambda(\mu)|$. Now to leading order in
powers of $\mu$ the $\beta$-functions for $\lambda$ are given by 
$(4\pi)^2\,\beta^{(1)}_\lambda(\mu)=2\,\lambda^3\,(\mu/M_c)+\cdots$, and
$(4\pi)^4\,\beta^{(2)}_\lambda(\mu)=-2\,\lambda^5\,(\mu/M_c)^2+\cdots$
~\cite{manel,zurab2}. 
In this case, the condition
$|\beta^{(2)}_\lambda(M_U)|\simeq|\beta^{(1)}_\lambda(M_U)|$, 
for the perturbation theory to be trustable, occurs for
$\lambda(M_U)\simeq 4\pi \sqrt{M_c/M_U}$, that we are using as
boundary condition to obtain the Higgs mass bound.
\begin{figure}[ht]
\centering
\epsfig{file=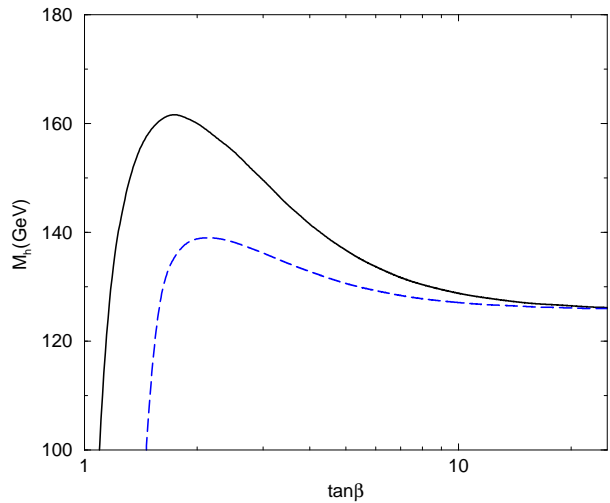,width=0.5\linewidth}
\caption{Upper bounds on $M_H$ for the 5D (solid line) and  4D (dashed
line) NMSSM.}
\label{figure}
\end{figure}
The result of the numerical analysis is shown in Fig.~\ref{figure}, where
we show the absolute upper bound on the lightest Higgs mass in the
unifying model above (solid line). We have fixed, for all values
of $\tan\beta$, $M_c=4$ TeV, which is consistent with indirect
bounds on $M_c$ from electroweak precision measurements.
We compare it with the
corresponding curve for the 4D NMSSM unifying at $M_{U}^{4D}$ and
with logarithmic running of the couplings. To obtain the curves of 
Fig.~\ref{figure} we have used initial conditions for the gauge
couplings, $h_t$ and $h_b$, according to the corresponding
experimental values for gauge couplings 
and the top and bottom-quark masses. Radiative
corrections are taken from the MSSM, with $M_{\rm SUSY}=1$ TeV and
maximal stop mixing. 

If one of the Higgs fields, $H_1$ ($H_2$), is in the bulk and the 
other, $H_2$ ($H_1$), in the 4D
boundary, one can still have the coupling (\ref{superp}) consistent
with Eq.~(\ref{condition}) if the singlet(s) $\vec{S}$ is
in the 4D boundary. However we will not consider this case in detail
since the obtained bound is weaker than in the previous one. In fact
using (\ref{wavedef}) and (\ref{wave}) one can see that now
$\tilde{b}_\lambda=3$ and, assuming
as before a model with $\rho=1$, we get $r=4/3$ that yields, using
(\ref{upperapp}), an absolute upper bound $\sim 147$~GeV.

Up to this point we have considered a genuine 5D mechanism by which
the lightest Higgs mass upper bound in the NMSSM 
unifying at a given scale $M_U$ can be
enhanced, i.e. the  
effect by which $\tilde{b}_\lambda/b_{\lambda}<1$. There is also another
possibility, in particular $\rho<1$, that we have
not yet considered. However this possibility, already existing in 4D
models unifying at scales $\ll M_{U}^{4D}$~\cite{espinosa1}, 
will not be discussed  in
great length. We only wish to mention how it can be
achieved in the class of the considered 5D models. Assuming $H_1$ and
$H_2$ in the 4D boundary Eq.~(\ref{unif}) can be satisfied for models
with $\rho<1$. For instance, considering 
right-handed quarks and leptons living in the bulk, $\Delta
N_E=2$ (two extra $E$-like hypermultiplets in the bulk) $T_1=1$ and
$T_0=1/2$ (one hypermultiplet in the $(\mathbf{1},\mathbf{3})_0$), one
obtains a model that unifies at one-loop as the MSSM but with
$\rho=1/2$. In this case $r=4$ and, using Eq.~(\ref{upperapp}), one
gets an absolute upper bound on $M_H$ as $\sim$~208 GeV. However, as discussed
earlier, in this class of models the condition for the dominance of the
KK threshold effects $M_c\ll M_s$ can be endangered if
$\rho\ll 1$.

A further interesting point is how robust is the present bound on the
Higgs mass, shown in Fig.~\ref{figure}, with respect to the number of
dimensions $\delta$~\footnote{We are assuming here that the heavy KK
excitations are arranged into $N=2$ 4D supermultiplets. Our results
crucially depend on this assumption, which is naturally realized in
models with $\delta=1,2$.} and also with respect to 
the regularization scheme used in
Eqs.~(\ref{RGE}) through (\ref{RGE2}). For an arbitrary number of
dimensions $\delta$ and regularization scheme choice, we should replace in
Eqs.~(\ref{RGE}) through (\ref{RGE2}), $e^t \to f_\delta e^{\delta\,
t}$, and in (\ref{escala}), $M_U/M_c \to \left(f_\delta/\delta\right)
\left(M_U/M_c\right)^{\delta}$, where the function $f_\delta$ depends
on the chosen regularization scheme and orbifold
projection. In this way the $\beta$-function
for $\lambda$ behaves as $16\pi^2\beta_\lambda=\widetilde{b}_\lambda
\lambda^3 f_\delta e^{\delta t}+\cdots$, the running is faster than
linear if $\delta>1$ and also depends on the regularization scheme
$f_\delta$. However using the resulting 
modified relation between $M_U^{4D}$ and
$M_U$, 
$$\frac{f_\delta}{\delta}\left(\frac{M_U}{M_c}\right)^\delta=
\rho\log\frac{M_U^{4D}}{M_c}\, ,
$$
we can see that one obtains the approximate relation at low
energy, $\lambda^2\simeq r\, \lambda_{4D}^2$, where $r$ is given in
(\ref{erre}), and so the relation (\ref{upperapp}),
involving the lightest Higgs mass $M_H$, holds. In other words, a change in
the number of dimensions $\delta$ and/or the renormalization scheme
$f_\delta$, that can modify the rate of running of all parameters of
the theory, can be approximately encoded in a change of the value of
the unification scale $M_U$, while the prediction on the upper bound of the
lightest Higgs mass remains invariant.   

In summary, we have considered how the presence of extra dimensions
can change the absolute upper bound on the lightest Higgs mass in
extensions of the MSSM with singlets. In particular we have found that
if either of the Higgs or singlet multiplets live in the bulk of the
extra dimension, because of the absence of wave function renormalization of
$N=2$ hypermultiplets to leading order, the $\beta$-function of the
Yukawa coupling
involving the singlet and Higgs superfields is suppressed, leading to
a slower running of the coupling and to a larger value at low scale,
with the corresponding enhancement of the tree-level contribution to
the Higgs mass. This effect raises the Higgs mass from $\sim 140$ GeV
in 4D models to $\sim 165$ GeV in models with extra dimensions and
some matter fields living in the bulk. The low
energy behaviour of these models is like the NMSSM but with an
enhanced lightest Higgs mass. This enhancement corresponds to the
range of energies that will be explored in the near future and can
thus have an important phenomenological impact. 
 
\newpage

\section*{Acknowledgements} 
The work of AD was supported by the Spanish Education Office (MEC)
under an \emph{FPI} scholarship. We acknowledge usefull discussions with 
C. Kounnas and F. Zwirner. AD also wants to thank the 
LPT of Ecole Normale Sup\'erieure for hospitality.

\end{document}